\documentclass[10pt,
               rmp,          
               showkeys,
               twocolumn,
              ]{revtex4}
\usepackage{graphicx}
\usepackage{amsmath,amssymb}

\newcommand{\spliter}         
{\vskip1ex
 \centerline{\rule{0.6\columnwidth}{0.1mm}}
 \vskip1ex}

\begin{document}
\title{Stickiness in mushroom billiards}
\author{Eduardo G. Altmann}
\email[Author to whom correspondence should be sent. E-mail address:
      ]{edugalt@mpipks-dresden.mpg.de}
\affiliation{Max Planck Institute for the Physics of Complex Systems, N\"othnitzer
  Strasse 38, 01187 Dresden, Germany} 

\author{Adilson E. Motter}
\affiliation{Center for Nonlinear Studies and Complex Systems Group,
  Theoretical Division, Los Alamos National Laboratory, MS B258, Los Alamos, NM 87545, USA}
\affiliation{Max Planck Institute for the Physics of Complex Systems, N\"othnitzer Strasse 38, 01187 Dresden, Germany} 

\author{Holger Kantz}
\affiliation{Max Planck Institute for the Physics of Complex Systems, N\"othnitzer
  Strasse 38, 01187 Dresden, Germany.} 

\date{\today}

\begin{abstract}
We investigate dynamical properties of chaotic trajectories in 
mushroom billiards. These billiards present a
well-defined simple border between a single regular region and a single chaotic component. We 
find that the stickiness of chaotic trajectories near the border of 
  the regular region
occurs through an infinite number of marginally
unstable periodic orbits. These orbits have zero
measure, thus not affecting the ergodicity of the chaotic
region. Notwithstanding, they govern 
the main dynamical properties of the system. In particular, we show that the marginally unstable periodic orbits explain the
periodicity and the power-law behavior with exponent~$\gamma=2$ observed
in the distribution of recurrence times.
\end{abstract}


 \maketitle

{\bf \noindent
The stickiness of chaotic trajectories in Hamiltonian systems, characterized
by long tails in the recurrence time statistics, is usually associated with the
presence of
partial barriers to the transport in the neighborhood of hierarchies of
Kolmogorov-Arnold-Moser (KAM) islands. However, as we show, these hierarchical 
structures are not necessary for the occurrence of stickiness. Here we study mushroom
billiards~\cite{bunimovich.mushroom}, which are analytically solvable
systems without hierarchies of KAM islands, and we show that these systems
present stickiness due to the presence of one-parameter families of marginally unstable
periodic orbits within the chaotic region. 

\spliter}


\section{\label{sec.I} Introduction}

An important property of the dynamics in
Hamiltonian systems with mixed phase space, where regions of regular and
chaotic motion coexist, is the
stickiness~\cite{karney} of chaotic trajectories  near the border of KAM islands.
These chaotic trajectories experience long periods of
almost regular motion, which strongly influence global
properties of the system, such as  transport~\cite{zaslavsky} and decay of
correlations~\cite{karney}. Even though no universal scenario exists~\cite{zaslavsky}, 
stickiness in Hamiltonian systems is in general related to topological properties
of invariant structures in the phase space~\cite{zas.trap,ketzmerick}. In the widely
studied case of Hamiltonian systems with hierarchies of infinitely many KAM
islands, there has been an intense debate about the possible
universality of the scaling exponents in the recurrence time statistics
(RTS)~\cite{meiss.ott.prl,chirikov,zaslavsky,ketzmerick,grassberger}. 
The difficulty in that case is partially due to the complexity of the invariant
structures in the phase space. It is thus of substantial interest the study of
Hamiltonian systems with simpler phase space, where the RTS can be studied analytically.

\begin{figure}[!ht]
\centerline{
\includegraphics[width=\columnwidth]{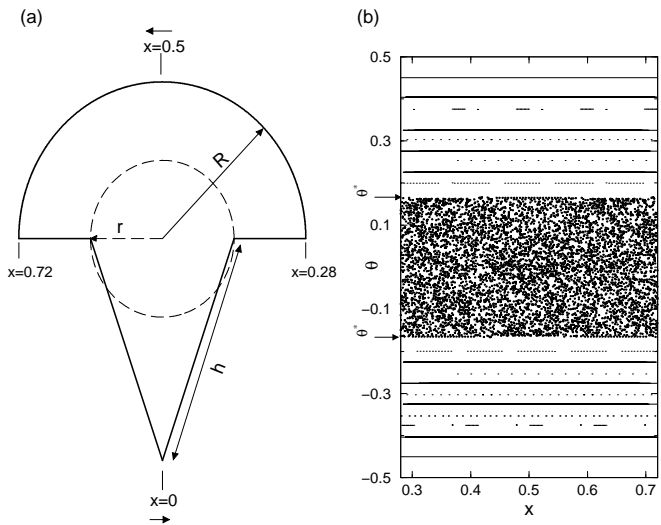}}
\caption{A mushroom billiard with triangular foot. (a) Configuration space.
(b) Phase-space representation of the semi-circular hat,
  where $x$ is the normalized position and $\theta$ is the normalized
  reflection angle. 
  The parameters are $r/R=0.6125$ and $h/R=1.5$, what
  implies~$\theta^*=\pm 1/6$. } 
\label{fig.spaces}
\end{figure}

Recently, Bunimovich~\cite{bunimovich.mushroom,bunimovich.lab} introduced a
new family of
Hamiltonian systems, the so called mushroom billiards, which have the
remarkable non-generic property of having a phase space with a single
KAM island and a single ergodic chaotic region. In
FIG.~\ref{fig.spaces}, we show  an example of mushroom billiard consisting
of a semi-circle ({\it hat}) placed on top of a  triangle ({\it foot}). 
In the configuration space, the border between the island and the chaotic sea consists of the orbits
in the semi-circular hat of the billiard that are tangent to  
the circle of radius~$r$ in FIG.~\ref{fig.spaces}a. The regular region (KAM
island) is composed of the trajectories that do not cross this circle, remaining forever 
in the semi-circular hat. The chaotic component
consists of the complementary set of trajectories, those 
that cross the dashed circle and may visit the foot of the mushroom. 
Except for zero measure sets, which are discussed in detail
  in this article, all trajectories in the chaotic region indeed visit the
  foot of the mushroom.
The stickiness close to the border of the chaotic region does not depend on
the specific format of the foot as far as
the foot provides a chaotic injection of the trajectories into the hat (where the
stickiness occurs). 
The coordinates used to describe the phase space are the normalized position~$x\in[0,1]$ along
the boundary of the billiard and the normalized angle~$\theta\in[-0.5,0.5]$ with respect to the normal
vector right after the specular reflection. With respect to these coordinates, the border
between the regular and the chaotic regions for trajectories in the semi-circular hat is at
$\theta^*(r/R) = \pm \sin^{-1}(r/R)$, as shown in Fig.~\ref{fig.spaces}b. The important control 
parameter is the ratio~$r/R$ and 
the time is counted as the number of reflections.  
Precise experimental realizations of billiards, e.g., in microwave
cavities and atom optics~\cite{experimental}, have generated additional
motivation for the investigation of the dynamics of these billiards.

%
 
In this article, we study the properties of stickiness in mushroom
billiards. Two important features are observed in the RTS: ({\it i}) 
periodicities in the sequence of recurrence times and ({\it ii}) a
power-law distribution with scaling exponent~$\gamma=2$ in the limit of long recurrence
times. Both effects are consequence of the presence
of one-parameter families of marginally unstable periodic orbits (MUPOs) inside the chaotic
region, which are studied here in detail. While the specific properties of the former effect are strongly
dependent on the particular value of the control parameter~$r/R$, the latter is invariant under  
changes of~$r/R$. 

The article is organized as follows. In Sec.~\ref{sec.numerical}, we
motivate the problem with numerical observations of stickiness in mushroom
billiards. The general characterization of MUPOs in mushroom billiards is presented in
Sec.~\ref{sec.mupos}. In Sec.~\ref{sec.teoria}, MUPOs are used to
describe the effects~({\it i}) and~({\it ii}) in the stickiness of the chaotic
trajectories. A summary of the conclusions is presented in the last section. 


\section{Numerical observation of stickiness}\label{sec.numerical}

In this section, numerical observations of the stickiness in
mushroom billiards are reported. We quantify the stickiness through the
RTS, as follows. The whole foot of the billiard is chosen as the recurrence
region. A typical trajectory is initialized within the recurrence region and followed  for a long
time. The time~$T$ the trajectory takes, after leaving the 
recurrence region, 
to return to it for the first time is recorded. Through further iterations of
the same trajectory, an
arbitrarily long sequence of recurrence times, distributed according
to~$P(T)$, can be obtained. The RTS is then defined as
\begin{equation}\label{eq.Q}
Q(\tau)= \sum_{T=\tau}^{\infty} P(T) = \lim_{N\rightarrow \infty} \frac{N_\tau}{N}\;,
\end{equation}
where $N$ is the total number of recurrences and $N_\tau$ is the number of
recurrences with time~$T\geq\tau$. For long times, in hyperbolic chaotic systems, the RTS decays 
exponentially~\cite{altmann}, while in systems with 
stickiness, it decays roughly as a power law~\cite{karney,zaslavsky},
\begin{equation*}\label{eq.powerlaw}
Q(\tau) \propto \tau^{-\gamma},
\end{equation*}
where~$\gamma>1$ is the scaling exponent~(see also Ref.~\cite{motter}). In our numerics, we approximate
Eq.~(\ref{eq.Q}) with a finite, statistically significant number~$N$ of recurrences.

\begin{figure}[!ht]
\centerline{
\includegraphics[width=\columnwidth]{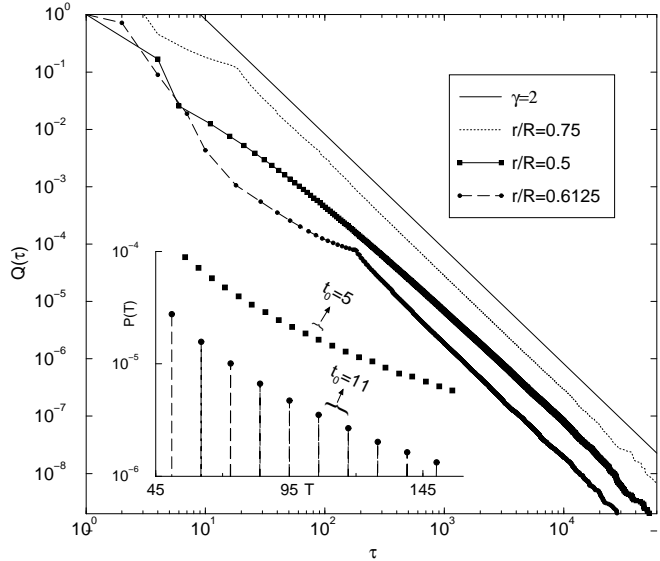}}
\caption{RTS of the mushroom billiard for various choices of the control
  parameter~$r/R$. The results are consistent with a power-law tail with
  exponent~$\gamma=2$. The distribution for~$r/R=0.75$ was shifted
  vertically upward by one decade for clarity. Inset: the distribution~$P(T)$ in the
  interval~$50 < T< 150 $ for~$r/R=0.5$ ($t_0=5$) and~$r/R=0.6125$
  ($t_0=11$).} 
\label{fig.ret}
\end{figure}

Our main observations about the RTS in the mushroom billiard are illustrated in
FIG.~\ref{fig.ret} and can be summarized as follows:

\begin{itemize}
\item [({\it i})] The recurrence times~$T$ for which recurrences
  are observed appear in a very organized way: times without a single
  recurrence~($P(T)=0$) are periodically interrupted by times with a high
  recurrence time probability. The period~$t_0$ between successive times with
  positive probability ($P(T)>0$) strongly depends on the control
  parameter~$r/R$ and may change over large intervals of time~$T$. In
  particular, as shown in the inset of FIG.~\ref{fig.ret} for recurrence
  times in the interval~$50 < T< 150 $, this period is~$t_0=5$ for $r/R=0.5$ and $t_0=11$ for
  $r/R=0.6125$. For longer recurrence times, higher periods may coexist
  with short periods.

\item [({\it ii})] The overall behavior of the RTS~$Q(\tau)$ shown in
  FIG.~\ref{fig.ret} presents a clear power-law tail with
  exponent $\gamma=2$, independently of the parameter~$r/R$.  
\end{itemize}

These observations are fully explained through the analysis of MUPOs present
in the chaotic component of the billiard. The existence and properties of
these orbits are discussed in the next section.


\section{MUPOs inside the chaotic region}\label{sec.mupos}

The chaotic trajectories in the mushroom billiard always visit the 
foot of the mushroom while the regular trajectories are confined to the
hat of the billiard. The integrability of the latter trajectories is based on the
conservation of the reflection angle~$\theta$ for collisions in 
the semi-circular hat. The novel aspect shown here is that there are
angle-preserving periodic orbits {\it inside} the chaotic region that never
visit the foot of the mushroom, and they form one-parameter families of
MUPOs. We use the acronym MUPO to refer specifically to
marginally unstable periodic orbits belonging to this class of orbits. They 
have zero Lyapunov exponent and, in contrast with elliptic points, their
eigenvalues are real and have modulus one. 
In this section, we study the stability and distribution of these MUPOs based
on the analysis of an equivalent open circular billiard.

\subsection{Open circular billiard}\label{ssec.circular}

\begin{figure}[!ht]
\centerline{
\includegraphics[width=\columnwidth]{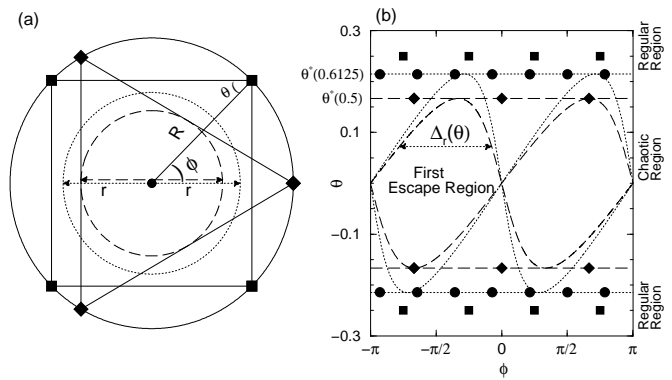}}
\caption{(a) The open circular billiard for two different lengths of the hole:~$r/R=0.5$ and
  $r/R=0.6125$. Diamonds~({\small$\blacklozenge$}) correspond to a periodic orbit~$(q=3,\eta=1)$
  and squares~({\tiny$\blacksquare$}) to a periodic orbit~$(q=4,\eta=1)$. (b)
  Phase-space representation of~(a). Circles ($\bullet$) correspond to a
  periodic orbit $(q=7,\eta=2)$, which is studied in detail in
  FIG.~\ref{fig.escape}, and the other symbols are the same as in~(a). The four
  horizontal lines represent 
  the border~$\theta^*$ between the chaotic and regular regions. The first
  escape regions for~$r/R=0.5$ and~$r/R=0.6125$ are the 
  areas limited by the dashed and dotted curves, respectively.}
\label{fig.circular}
\end{figure}

\begin{figure}[!ht]
\centerline{
\includegraphics[
width=\columnwidth
]{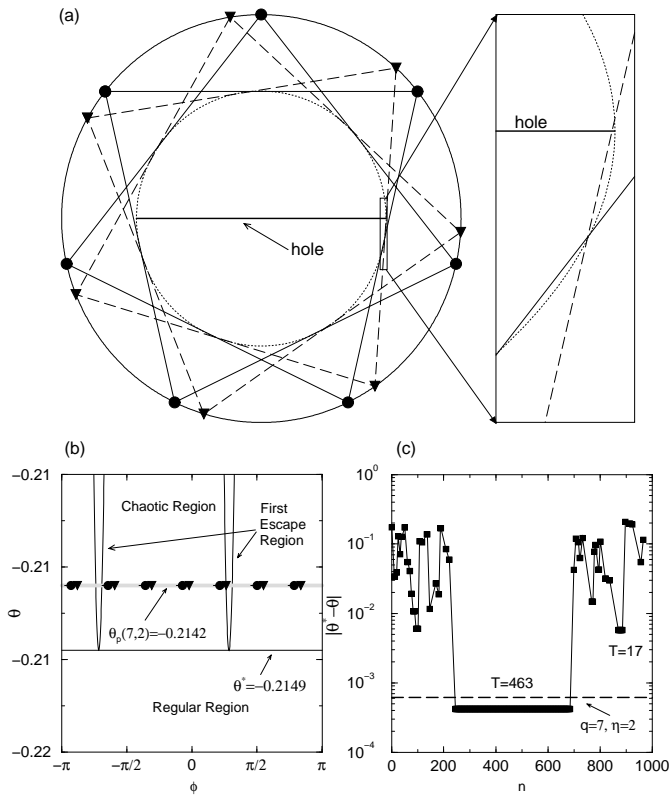}}
\caption{Detailed analysis of orbits~($q=7,\eta=2$) for 
  $r/R=0.6125$. (a) Configuration space, where two orbits ($q=7,\eta=2$) that cross the circle
  with radius~$r$ are shown. The orbit
  represented by circles ($\bullet$) does not hit the hole, while the orbit
  represented by triangles~($\blacktriangledown$) hits the hole on the right-hand side
  (see amplification). (b) Phase-space representation of the
  orbits in (a), where it is shown that they are respectively outside
  and inside the first escape region. A small perturbation 
  in the reflection angle~$\theta$ of the first orbit leads to a continuous rotation of the orbit in (a) 
  [horizontal drift in (b)] until the trajectory hits the hole [enters the first escape
  region in (b)]. (c) Time evolution of the distance from the regular
  region to a chaotic trajectory that approaches the family of MUPOs~($q=7,\eta=2$) in
  the original mushroom billiard. Events with recurrence time $T=463$ and
  $T=17$ are highlighted. The events with large recurrence times are associated with
  approaches to the MUPOs.}
\label{fig.escape}
\end{figure}

 A convenient way to
visualize the MUPOs is to consider a circular billiard of radius~$R$, as depicted in
FIG.~\ref{fig.circular}a, which has the property that trajectories are
considered to escape
when they hit the horizontal straight-line segment of 
length~$2r$  in the center of the billiard (hereafter referred to as the {\it
  hole}). The equivalence between the two billiards is based on the application of the image
construction ``trick'' to the horizontal segments
in the hat of the mushroom billiard and on the independence on the shape of the foot~\cite{bunimovich.mushroom}.
The coordinates of the circular billiard   
are the reflection angle~$\theta\in[-0.5,0.5]$ with respect to the normal
vector and the position of collision in the circumference, given by the
angle~$\phi\in[-\pi,\pi]$, as indicated in
FIG.~\ref{fig.circular}a. The time is again measured as the number of
  reflections at the border of the billiard. This introduces a minor
  difference between the two billiards since in the mushroom billiard one
  counts the reflections at the horizontal segments of the hat. Nevertheless,
the dynamics of the trajectories in  
the open circular billiard is equivalent to that of trajectories in the semi-circular hat of the mushroom
billiard, where stickiness occurs and where the MUPOs are located. Geometrically, the MUPOs are the periodic
orbits of the open circular billiard that cross the circle of radius~$r$ but 
that do not hit the hole in the center of this circle. Examples of MUPOs are
shown in Figs.~\ref{fig.circular} and~\ref{fig.escape} for the
parameters~$r/R=0.5$ and~$r/R=0.6125$. 

We now describe the necessary and sufficient conditions for the existence  of
MUPOs inside the chaotic region of the mushroom billiard. A periodic
orbit in the open circular billiard is defined by two integer numbers: the period~$q$ and the
rotation number~$\eta$ (i.e., the number of laps around the center of the circle), where~$q/\eta$ is an
irreducible fraction and~$q>2\eta$. The invariant reflection angle~$\theta$ of
this orbit is given by~$\theta_p(q,\eta)=\pm(q-2\eta)/2q$. 

The first necessary 
condition for the existence of a MUPO defined by~$q$ and $\eta$ is that
the corresponding trajectory crosses the circle of radius~$r$,  
meaning that the trajectory would be
inside the chaotic component of the original mushroom
billiard. Accordingly, the reflection angle of this orbit must satisfy
\begin{equation}\label{eq.apendice}
|\theta_p(q,\eta)|=\frac{q-2\eta}{2q} < \frac{1}{\pi}\sin^{-1}(r/R)=|\theta^*(r/R)|.
\end{equation}
The second necessary condition for this orbit to be a MUPO is that
its trajectory does not hit the hole. In 
order to study this condition in the phase space of the open circular billiard, we
analyze the points with reflection angle~$\theta$ that hit the hole in
one time step. These points define what we call the {\it first} escape region, whose
width we denote by~$\Delta_r(\theta)$ (see
FIG.~\ref{fig.circular}b). The trajectory does not hit the hole if all the~$q$ 
periodic points of the orbit are outside the first escape region.  With respect to
the coordinate~$\phi$, the distance between two neighboring periodic points is constant,
namely  $2\pi/q$. The second condition can thus be written as 
\begin{equation}\label{eq.delta}
\frac{2\pi}{q}> j \Delta_r(\theta_p),\;\;\; j=\left\{
\begin{array}{ll} 1 & \mbox{if $q$ is even,} \\
                  2 & \mbox{if $q$ is odd.}
\end{array}
\right .
\end{equation}
The factor~$j=2$ for the odd-period periodic orbits comes from
the~$2\pi$ periodicity of the points of these orbits in opposition to the~$\pi$ periodicity
of the escape region. 

To calculate $\Delta_r(\theta)$, we have to determine the borders of the first
escape region (FIG.~\ref{fig.circular}b). For a given 
position~$\phi$ in the circumference of radius~$R$, the angles~$\theta$ of
the trajectories that first hit the hole are limited by the angles~$\theta^\pm$ given by
\begin{equation}
\begin{array}{ll}
\theta^+&= \frac{1}{2\pi}[\phi+\tan^{-1}(\frac{R\sin(\phi)}{r+R\cos(\phi)})], \\\\
\theta^-&=\frac{1}{2\pi}[\frac{\pi}{2}-\phi+ \tan^{-1}(\frac{R \cos(\phi)-r}{R\sin(\phi)})].
\end{array}
\end{equation}
The width of the first escape region is then given by 
\begin{equation}\label{eq.D}
\Delta_r(\theta_p)=\phi^{\pm}_1-\phi^{\pm}_2\;,
\end{equation}
where~$\phi_1^{\pm}>\phi_2^{\pm}$ are the two solutions of the
equation~$\theta^{\pm}=\theta_p$. Observe that, because of the
symmetry~$\theta\rightarrow -\theta$, we have
$\phi_1^+-\phi_2^+=\phi_1^--\phi_2^-$.

\begin{figure}[!ht]
\centerline{
\includegraphics[width=\columnwidth]{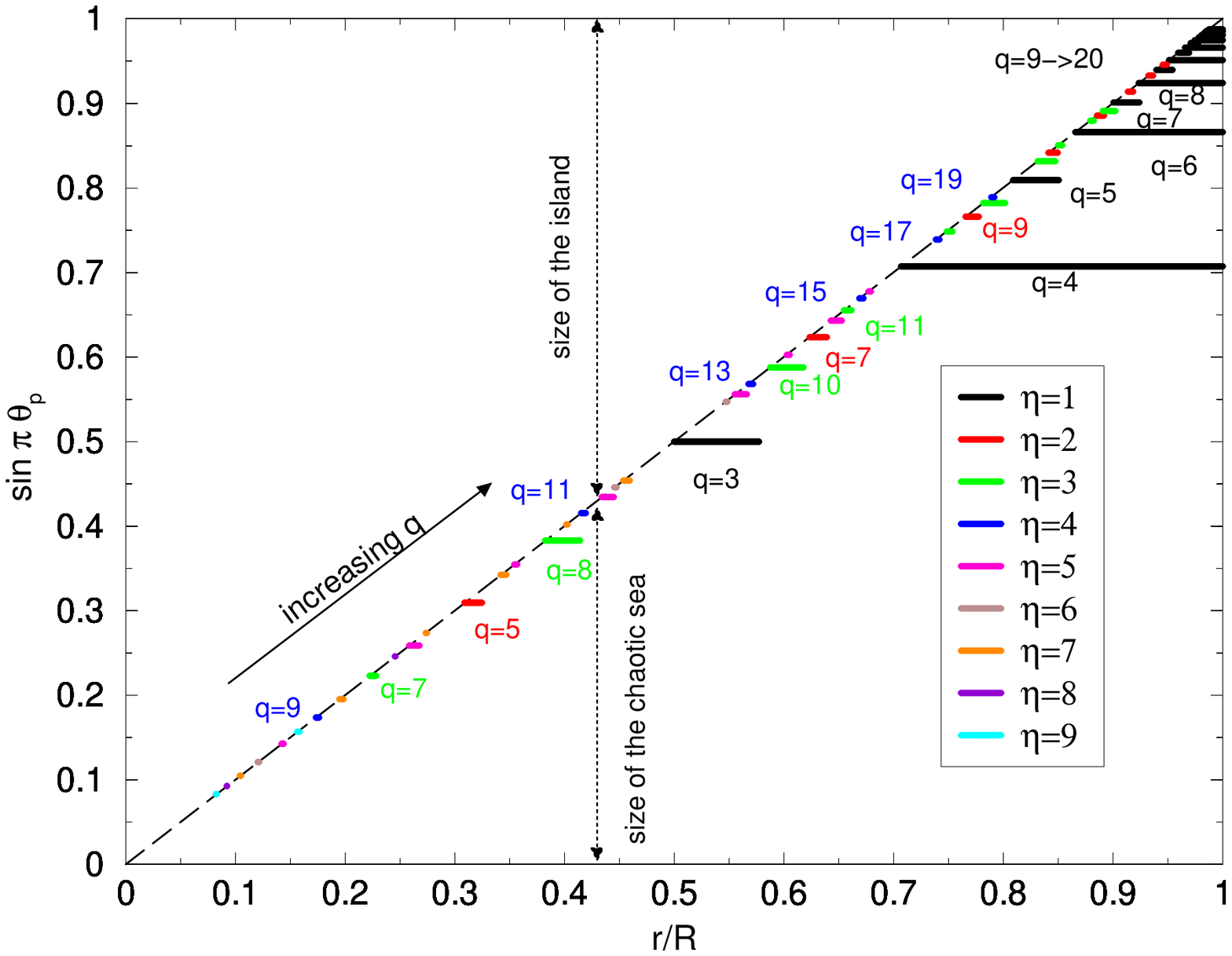}}
\centerline{
\includegraphics[width=\columnwidth]{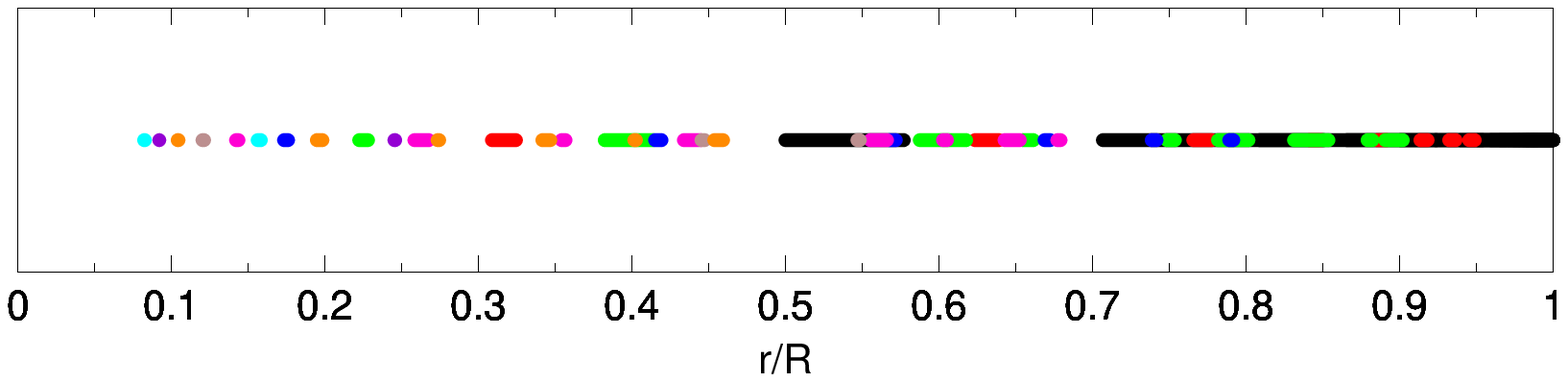}
}
\caption{(Color) Intervals of the control parameter $r/R$ for which orbits
  $(q,\eta)$ are MUPOs. All orbits with $q\leq 20$ are shown.}
\label{fig.mupos}
\end{figure}

Gathering all these, we have that the conditions expressed in the
Eqs.~(\ref{eq.apendice}) and (\ref{eq.delta})  are not only
necessary but also sufficient for an orbit with period~$q$ and
rotation number~$\eta$ to be a MUPO inside the chaotic region of the original
mushroom billiard. 
These conditions can be translated in terms of the interval of the control 
parameter $r/R$ for which a given periodic orbit $(q,\eta)$ is a MUPO: 
\begin{equation}\label{eq.mupos}
  \sin \left[ \pi \theta_p(q,\eta) \right]
    < r/R < \frac{\sin \left[ \pi
  \theta_p(q,\eta)\right] }{\cos \left[ \pi/(j q)\right] }\;,
\end{equation}
with $j$ as in Eq.~(\ref{eq.delta}).
In Fig.~\ref{fig.mupos}, we show the parameters for which the orbits up to
period~$q=20$ are MUPOs. An efficient procedure to find higher order MUPOs for
a given parameter $r/R$ is to take $\eta/q$ as the convergents of 
the continuous fraction expansion of $\frac{1}{\pi} \cos^{-1}(r/R)$ and
verify if they fulfill condition~(\ref{eq.mupos}) or, equivalently,
Eqs.~(\ref{eq.apendice}) and~(\ref{eq.delta}). Through this procedure we have found
the following MUPOs for $r/R=0.625$: $(q,\eta) =$ 
(7, 2), (698, 199), (1161, 331), (18341, 5229), (2136146, 609013), and (8526243, 2430823).

\subsection{Stability of periodic orbits}\label{ssec.stability}

In this section we show that the periodic orbits considered in
Sec.~\ref{ssec.circular} are  marginally unstable, i.e., are unstable but
have zero Lyapunov exponent, and that they form one-parameter families in the phase
space. Consider small perturbations to one of these orbits. If 
the initial position~$\phi$ is perturbed, neglecting the presence of the hole,
an equivalent orbit rotated in the configuration space is always obtained
(FIG.~\ref{fig.escape}).  When we consider the presence of the hole, some of
these orbits escape and create gaps in the possible~$\phi$ values of the periodic
orbits. However, when the (strict) inequalities in Eqs.~(\ref{eq.apendice})
and~(\ref{eq.delta}) are satisfied, there remain (open) intervals of~$\phi$
values in between these gaps. These remaining intervals define a one-parameter family of
MUPOs with constant~$\theta$.  
A small perturbation in the
angle~$\theta'=\theta-\varepsilon$, on the other hand, leads to an
increase~$\phi'-\phi=2\varepsilon$ per time step in the distance from the
original 
unperturbed orbit. This shows that the separation increases linearly in time and
hence that these periodic orbits are indeed marginally
unstable. Since the marginal stability does not depend on the
  discretization of time, the statistical properties of the recurrence time
  discussed in this article remain valid for continuous time dynamics.

Typically, a MUPO perturbed with respect to~$(\theta,\phi)$ is a
quasi-periodic orbit in the circular billiard, which, without the hole,
would  pass arbitrarily close to all values of~$\phi$. Because of this ergodicity
with respect to the coordinate~$\phi$, the corresponding perturbed trajectory eventually enters the first
escape region and hits the hole, as illustrated in FIG.~\ref{fig.escape}.  
The distance from the unperturbed MUPO grows linearly in time. However, the
distance~$\varepsilon$ from the corresponding {\it family} of MUPOs remain constant
until the perturbed trajectory hits the hole and escapes (FIG.~\ref{fig.escape}c).

\subsection{Distribution of families of MUPOs}\label{ssec.remains}

\begin{figure}[!ht]
\centerline{
\includegraphics[width=\columnwidth]{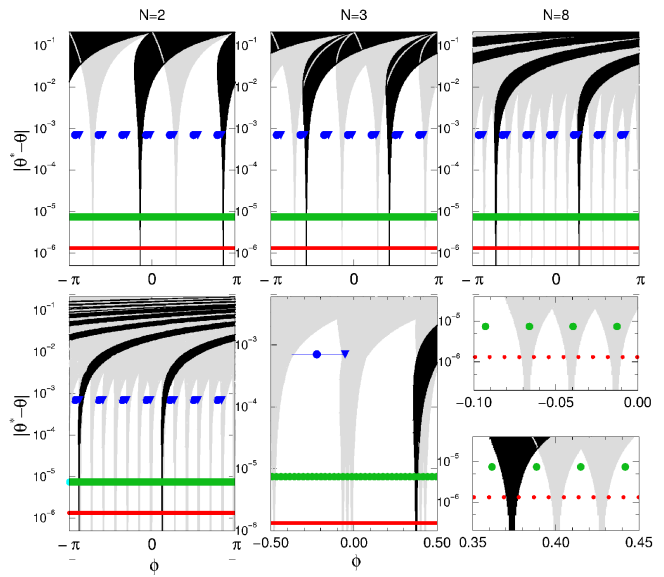}}
\caption{ (Color online) Escape regions in the the phase space of the
  open circular billiard for
  $r/R=0.6125$. In each panel, the $N$th escape region is shown in black and
   the $n$th escape regions for all~$n<N$ are shown in gray. In the first row, we show the
  cases~$N=2,N=3,$ and $N=8$, while in the second row we show the case~$N=20$
  and successive amplifications for this case. Different symbols correspond to the different orbits~$(q=7,\eta=2)$,
  $(q=235,\eta=67),$ and~$(q=698,\eta=199)$, from
  top to bottom, respectively. The figures at the bottom right
  show that only the first and the last of these orbits are MUPOs.}  
\label{fig.remains}
\end{figure}

We now use the information about the first escape region,
Eq.~(\ref{eq.D}), to study how the families of MUPOs are distributed inside
the chaotic region of the mushroom billiard. The MUPOs are the
only orbits in the mushroom's hat that cross the circle of radius~$r$ but 
never visit the foot of the mushroom~(Fig.~\ref{fig.spaces}a). 
Equivalently, in the open circular billiard, these are the orbits that cut the
circle of radius~$r$ but never escape~(Fig.~\ref{fig.circular}a).  
In FIG.~\ref{fig.remains}, we show the phase space of the open circular billiard
for trajectories~$|\theta|<|\theta^*|$, associated with the chaotic region of the 
mushroom billiard, and the $N$th escape region, defined by the ($N-1$)th
pre-image of the first escape region shown in FIG.~\ref{fig.circular}b. 
 We note that the larger the period of the orbit the closer it may
be to the border~$\theta^*$.
For example, for the control parameter~$r/R=0.6125$ considered in
FIG.~\ref{fig.remains}, the border is at $\theta^*=0.2149010...$ and the orbits $(q=7,\eta=2)$,
$(q=235,\eta=67)$ and $(q=698,\eta=199)$ highlighted in the figure are at
$\theta_p=0.2143857...$, $\theta_p=0.2148936...$ and $\theta_p=0.2148997...$,
respectively. The widths of the escape regions go to zero when 
the border is approached. In Fig.~\ref{fig.remains}, the MUPOs correspond to
the points that do not belong to any of the $n$th escape regions, for
all~$n<N$ in the limit of~$N\rightarrow\infty$. 
That is, taking the limit $N\rightarrow\infty$ in FIG.~\ref{fig.remains},
all the points outside the escape regions belong to MUPOs. 
Note that a complex distribution of families of MUPOs
may exist near the border of the island.

\section{\label{sec.teoria} MUPOs and stickiness}

In this section, we show how the MUPOs characterized in Sec.~\ref{sec.mupos}
affect the stickiness of chaotic trajectories in the mushroom billiard
and explain the numerical results reported in Sec.~\ref{sec.numerical}. 
Qualitatively, the stickiness of chaotic trajectories in mushroom billiards can be
described as follows. A chaotic trajectory bounces in the foot
of the mushroom until it enters the semi-circular hat. Depending on
the reflection angle~$\theta$, which is conserved for collisions 
in the hat, the trajectory may be close to a MUPO. 
In this case, it remains close to the family of this
MUPO a long time before falling again into the foot of the billiard
(see FIG.~\ref{fig.escape}c). In the phase space, these
 events with large recurrence time are observed as a continuous rotation of the trajectory around
the regular island, at a fixed distance from the island, until it enters the first escape region.  

We now show how the MUPOs explain quantitatively the two effects observed in the RTS  
  and anticipated in Sec.~\ref{sec.numerical}:

\noindent {\bf {\it (i)} Regularity in the distribution~$P(T)$.} When a
chaotic trajectory visits the hat of the billiard, the approach to as well as
the escape from the neighborhood of a family of MUPOs takes place in a single
time step, as illustrated in FIG.~\ref{fig.escape}c by a sharp
  transition before and after the interval of minimum distance to the island. Due to the 
injection of trajectories from the foot
of the mushroom billiard, the approach to the MUPOs (that requires
an angle~$\theta$ close to~$\theta_p$)  can only occur close to some of the~$q$ 
periodic points of the MUPO. Actually, it happens always near the same
position in the mushroom's hat, right above the intersection point between
the boundary of the foot and the bottom part of the hat (i.e., one of the
two horizontal lines). Similarly, the escape occurs always when the trajectory is close to the 
point of the MUPO that lies in the bottom part of the mushroom's hat. Due to these constraints,
the intervals of time a chaotic trajectory that approaches a family of MUPOs spends away from the
mushroom's foot form a sequence that can be 
written as~$T_i=a-1+(q+2\eta) i$, where $i\in I\!\!N$ and $a$ is the time between
the first collision in the semi-circular hat and the 
collision in the bottom part of the hat (close to the hole). The period~$t_0$
between successive recurrence times with positive
probability, reported in Sec.~\ref{sec.numerical}, is thus related to 
the period~$q$ and the rotation number~$\eta$ of the MUPO inside the chaotic
region according to
\begin{equation}\label{eq.periodo}
               t_0 = q + 2\eta \;.
\end{equation}
The factor $2\eta$ comes from the fact that, in contrast with the
open circular billiard, in the mushroom billiard one counts the collisions in the
bottom of the hat. In FIG.~\ref{fig.circular} we show that for $r/R=0.5$ the
orbit~($q=3,\eta=1$) is exactly in the border between the regular
  island and the chaotic region. Through
Eq.~(\ref{eq.periodo}) we obtain~$t_0=5$, explaining the numerical
observation. Analogously for the parameter~$r/R=0.6125$, in FIG.~\ref{fig.escape} we see
the orbit~($q=7,\eta=2$) that implies~$t_0=11$, exactly as observed
numerically. Higher-order periodicities  are associated with the existence
of additional families of MUPOs in the chaotic sea, as shown in
Sec.~\ref{ssec.remains}. 

\noindent {\bf ({\it ii}) $\gamma=2$ in the tail of the RTS.}
In the stability considerations described in Sec.~\ref{ssec.stability}, 
we noticed that when we perturb a MUPO, the distance from the
unperturbed orbit increases linearly along~$\phi$ and remains
constant with respect to the reflection angle~$\theta$. 
These properties of the MUPOs in mushroom billiards are 
equivalent to those observed for MUPOs present in the much simpler case of
billiards with parallel walls. In the latter case, the MUPOs are
generated by trajectories bouncing perpendicularly between parallel
walls. Billiards with parallel walls, such as the Sinai billiard and  
stadium billiard, have been studied extensively~\cite{primeiros,gaspard,armstead,vivaldi}. It has been shown that
the existence of a one-parameter family of MUPOs due to parallel walls 
leads to an exponent~$\gamma=2$ in the power-law decay of the
RTS~\cite{gaspard,armstead,nosso}. Each family of MUPOs in the mushroom
billiards is equivalent to a family of MUPOs generated by the presence of parallel walls.
The same is true for the family of periodic or quasi-periodic marginally
unstable orbits always present at the border between the regular and chaotic
region of the mushroom billiard.
In the mushroom billiard, more than one family of marginally unstable orbits may exist.  Since
each of them leads to the same asymptotic 
exponent~$\gamma=2$, irrespective of the control parameter~$r/R$, the overall
RTS will also have scaling exponent~$\gamma=2$. A similar argument can be
used to show that the same exponent governs the RTS of honey
mushrooms~\cite{bunimovich.mushroom}, which are systems composed by a finite
number of mushroom billiards.


\section{\label{sec.V} Conclusions}

Mushroom billiards~\cite{bunimovich.mushroom} provide an example of non-trivial
Hamiltonian systems with a relatively simple phase space. In contrast with
previously considered Hamiltonian systems, which have a hierarchy of
infinitely many regular islands, the mushroom billiards considered here have a
single island. We have shown that, in spite of the simplicity
of the regular region, the chaotic
component of the mushroom billiards contains a complex distribution of
marginally unstable periodic orbits (MUPOs). These
orbits have zero measure and do not affect the ergodicity of the
system. Nevertheless, the MUPOs are 
crucial for the understanding of important dynamical properties of the system, as
shown here through the analysis of the recurrence time statistics. In particular, we have shown that
these MUPOs lead to an exponent~$\gamma=2$ for the asymptotic
power-law decay of the recurrence time statistics. This result leads to the interesting question of
whether this scaling exponent is universal within the class of Hamiltonian
systems with finite number of regular islands~\cite{nosso} which, in turn, may
provide new insight into the problem of universality in Hamiltonian system in
general.

\begin{acknowledgments}
E.G.A. is supported by CAPES (Brazil) and DAAD (Germany). A.E.M. is supported
by the U.S. Department of Energy under Contract No. W-7405-ENG-36.
\end{acknowledgments}


\end{document}